\documentclass[preprintnumbers]{revtex4}
\usepackage{amsfonts}
\usepackage{eurosym}
\usepackage{amsmath}
\usepackage{amssymb}
\usepackage{graphicx}
\usepackage{color}

\begin{document}

\date{\today }
\date{\today }
\title{Excitations of N$_{2 }$ and O$_{2}$ molecules due to helium ion
impact and a polarization effect}
\author{M. Gochitashvili$^{1}$, R. Lomsadze$^{1}$, R. Ya. Kezerashvili$%
^{2,3} $, I. Noselidze$^{1}$, and M. Schulz$^{4}$}
\affiliation{\mbox{$^{1}$Tbilisi State University, 0179 Tbilisi, Georgia }\\
$^{2}$New York City College of Technology, The City University of New York,
Brooklyn, NY 11201, USA \\
$^{3}$The Graduate School and University Center, The City University of New
York, New York, NY 10016, USA\\
$^{4}$Missouri University of Science and Technology, Rolla, MO 65409, USA}

\begin{abstract}
We present an experimental study of the dissociative excitation in the
collision of helium ions with nitrogen and oxygen molecules for collision
energy of $0.7-10$ keV. Absolute emission cross sections are measured and
reported for most nitrogen and oxygen atomic and ionic lines in wide, vacuum
ultraviolet ($80-130$ nm) and visible ($380-800$ nm), spectral regions.
Remarkable similarities of the processes realized in He$^{+}+$N$_{2}$ and He$%
^{+}+$O$_{2}$ collision systems are observed. We present polarization
measurements for He$^{+}+$N$_{2}$ collision system.

The emission of excited dissociative products was detected using an improved
high-resolution optical spectroscopy method. This method incorporates the
retarding potential method and a high resolution electrostatic energy
analyzer to precisely measure the energy of incident particles and the
energy of dispersion. The improvement in the optics resolution allows us to
measure the cross section on the order of 10$^{-19}$ cm$^{2}$ or lower.
\end{abstract}

\maketitle

\section{Introduction}
\label{Intro}

Nitrogen and oxygen molecules are the main constituents of the atmosphere
and are simple species of great interest for interstellar medium \cite%
{1M,2M,3M,4M}. First, all nitrogen and oxygen molecules are important
constituents of the upper atmosphere of planets and play a crucial role in
atmospheric chemistry at mesospheric and thermosphere altitudes. Solar
radiation mainly results from solar flares, solar wind, coronal mass
ejections, and solar prominences and includes electromagnetic radiation and
energetic electrons, protons, and $\alpha -$particles. These basic
components of solar radiation interact with atmospheric nitrogen and oxygen
molecules: electromagnetic radiation with energies from a few tenths of eV
to hundreds of MeV, electrons and protons in the energy spectrum from a few
tenths of eV to hundreds of MeV and up to GeV, and helium ions with energies
up to 10 keV. Numerous processes of the formation of atoms or atomic ions in
which oxygen and nitrogen molecules act as targets, have attracted
considerable interest. These processes play a crucial role in atmospheric
chemistry at mesospheric and thermosphere altitudes. Ionic species are
present in the upper atmosphere of planets and govern ionosphere chemistry.
Atomic and molecular ions have also been detected in the comet tails. Among
the many inelastic processes involving oxygen ions, excitation, ionization,
and charge exchange are relevant to the low-temperature edge plasma region
of current thermonuclear fusion devices \cite{5M,6M}. Oxygen is also typical
impurity in almost all laboratory plasmas \cite{5M}.

The electromagnetic radiation, the corpuscular part of the solar radiation,
that is electrons, protons, and helium ions \cite{10m} with an energy
spectrum between a few tenths of eV and hundreds of MeV, are heading towards
the Earth, and interact with the atmosphere. The corpuscular part of solar
radiation is deflected by the Earth's magnetic field towards the poles and
is scattered and absorbed by atmospheric atoms and molecules, particularly
by nitrogen and oxygen. The accompanying ionization of various gases in the
upper atmosphere causes a luminous glow in the upper atmosphere, the
so-called phenomenon of - aurora. Although there have been investigations of
the main characteristics of the aurora, there is still a lack of
quantitative description of this phenomenon. Upon reaching the denser layers
of the atmosphere, electrons and helium ions participate in various
inelastic processes such as ionization, molecular excitation, and
charge-exchange reactions on atmospheric gases, especially nitrogen and
oxygen molecules. Spectral analysis of the aurora shows that ionized
nitrogen molecules can radiate in the visible, infrared, and ultraviolet
region.

Observations of the excitation of B $^{2}\Sigma _{u}^{+}$ and A $^{2}\Pi _{u}
$ band systems in the ionized nitrogen molecule N$_{2}^{+}$ indicate their
presence in the aurora and dayglow \cite{11m,12m,13m,14m,15m,16m}. These
bands appear in the spectra of polar auroras and contain information on the
collision processes in the upper atmosphere. Hence, during collisions,
vibrationally excited N$_{2}^{+}$ ions and their radiative decay are
accompanied by the creation of electronic ground X $^{2}\Sigma _{g}^{+}$
states.

The study of emission spectra provides an opportunity to determine the
concentration and energy distribution of particles entering the upper layer
of the atmosphere. To address this problem, it is necessary to determine
high-precision absolute cross-sections of various inelastic processes, such
as ionization, excitation, and charge-exchange. However, the determination
of the absolute cross-section, for example, of the excited B $^{2}\Sigma
_{u}^{+}$ band system is challenging. The number of experimental studies in
which this excited band system is measured is very limited, and these
studies are usually related to the processes of excitation through electron
collisions with nitrogen molecules \cite{17m,18m,19m,20m,21m}. However, in
the case of electron impact, the experimentally determined cross-section for
the formation of N$_{2}^{+}$ ions in the A$-$state is only known to be
within 50\% because measurements of excitation cross-sections involve
various challenges. In particular, the lifetime of nitrogen molecule ions in
the A $^{2}\Pi _{u}$ state is approximately 10$^{-5}$ s \cite{17m,18m}, and
during measurements the quenching of excited particles (the transfer of the
excited energy to other particles) is expected to occur.

The dissociation of highly excited molecular states deserves special
interest and is the subject of extensive research \cite%
{22m,23m,24m,25m,26m,27m,28m,29m,30m,31m}, including experimental
investigations that add new information. The products of the molecular
dissociation may remain in an excited state. Highly excited states may decay
through pre-dissociation or autoionization channels with the formation of a
neutral atom, electron, or ion \cite{23m,24m,25m,27m}. For example, neutral
fragments resulting from the dissociation of highly excited states were
reported in Refs. \cite{26m,28m,29m,30m,31m}. The absolute cross-section of
the luminescence due to the excited product of atomic oxygen dissociation in
the wavelength range of 97--131 nm was measured for photoionization \cite%
{32m}.

Polarization information is important for the accurate determination of
absolute and relative photon emission cross-sections. The results of the
polarization measurements also allowed some nontrivial conclusions related
to the spatial distribution of the electron cloud. A reliable experimental
determination of the polarization fraction, not only provides additional
information about the details of excitation cross-sections by determining
the relative populations of the degenerate magnetic sublevels, but also
enables a comparison of available experimental data with calculations of
total cross-sections.

Quantitatively, the polarization fraction can be analyzed in terms of the
alignment of orbital momentum sublevels. The cross-sections of the
population of magnetic sublevels provide detailed information on the
excitation mechanism. Owing to the different populations of magnetic
sublevels within a certain ($nl$) subshell, the radiation can be polarized
and, consequently, anisotropic.

Numerous studies have investigated the polarization of radiation in ion-atom
and ion-molecule processes \cite%
{35m,36m,37m,38m,39m,40m,41m,42m,43m,44m,45m,46m,47m}. Usually, in
polarization measurements, coincidences of the photon and scattered particle
are detected \cite{38m,39m,40m,41m,42m,43m}.

In the case of inelastic He$^{+}-$N$_{2}$ and He$^{+}-$O$_{2}$ collisions,
the radiation spectrum is multitudinous. Therefore, a monochromator with a
high resolving power ($\sim $ 0.2 nm) should be used. To amplify the optical
registration sensitivity in addition to the monochromator, broad bandpass
filters were used for isolating the optical lines. An anisotropic excitation
mechanism is common in astrophysical plasmas and is readily reproduced in a
laboratory environment \cite{49m}. Almost a century ago \cite{50m}, was
shown that spectral line emissions originating from atoms or ions excited by
particles whose velocity distribution is anisotropic, in general, are
polarized. If collisional excitation occurs by impact in a preferred
direction, the magnetic sublevels of the excited states can be populated
with non-statistical probabilities. When the state decays, the emitted
electromagnetic radiation becomes spatially anisotropic and partially
polarized \cite{51m}.

In Ref. \cite{52m}, the authors studied the polarization of the radiation
emitted from ions excited by an electron beam impact inside an electron-beam
ion trap. This demonstrates that the polarization of the emitted radiation
is especially important when measurements are made with spectrometers in
which the energy disperser is polarization-selective. Polarization-sensitive
measurements may also be used to detect resonance processes that are too
weak to be directly observed. Moreover, this provides information about the
magnetic sublevels that would normally remain hidden in simple energy
dispersive measurements.

Applying conservation of angular momentum allows the calculation of the
relative populations of the magnetic sublevels \cite{53m}. The magnetic
sublevel population of the autoionizing states of helium, excited by the
charged particle impact was determined in Refs. \cite{54m,55m}. The degree
of polarization of the radiation emitted by atoms and ions following
particle impact contains information on the excitation of magnetic sublevels
with different projections $M$ of orbital momentum \cite{56m}. In Ref. \cite%
{54m}, the authors measured the degree of linear polarization in the extreme
ultraviolet region and cross-sections of excitation to individual magnetic
sublevels. For the 1s-2p single-electron excitation of helium, it was found
that the effects are stronger for the excitation of sublevels with $M=0$,
than with $M=\pm 1$. In the literature, considerable evidence exists that
molecular radiation may be strongly polarized for both discharge sources and
when electron beam excitation is used \cite{61m}. In Ref. \cite{62m}, the
authors predicted that atomic or ionic radiation following dissociative
excitation of molecules can be polarized. Several excited states of the N
and N$^{+}$ were identified in the visible and near-infrared optical
emission spectra produced by the electron impact excitation of the N$_{2}$
(X $^{1}\Sigma _{g}^{+}$ ) \cite{63m}. During the dissociation process, one
of the fragments (ion/atom) can be left in an excited state from which it
radiates. The fluorescence from the excited atomic fragment can be
polarized, and the degree of this polarization is related to the form of
anisotropy in the angular distribution of the dissociation product \cite{64m}%
.

In the last few years, experimental studies on ion collisions at low and
medium energies (a few eV and keV) have been performed using different
techniques. For example, in the case of the interaction between He$^{+}$
ions and H$_{2}$, N$_{2}$, O$_{2}$, CO, and NO molecules, collision
spectroscopy methods \cite{33m,66m,67m}, and high resolution translational
spectroscopy for the pairs H$_{2}^{+}-$H$_{2}$, H$_{2}^{+}-$Mg, H$_{2}^{+}-$%
Na, H$_{2}^{+}-$Cs, and H$_{2}^{+}-$Ar \cite{68m} have been used. Emission
from the dissociation products in the visible range revealed a rich spectrum
of excited states in the collisions of different ions with O$_{2}$ and N$_{2}
$ molecules \cite{69m,70m}.

In this work, we experimentally studied dissociative excitation in
collisions of helium ions with nitrogen and oxygen molecules in the ion
energy range of 0.7--10 keV. One of the reasons for the choice of the helium
ion as a projectile is that highly excited molecular states of the oxygen
and nitrogen molecular ion arise in this case because the inelastic channels
of charge exchange prevail in the respective range of collision energies,
and thus the inner-shell electrons of the molecules are captured \cite%
{33m,34m}. Absolute emission cross-sections were measured and reported for
most nitrogen and oxygen atomic and ionic lines in the vacuum ultraviolet ($%
80-130$ nm) and visible ($380-800$ nm), spectral regions. The measurements
were performed using high-resolution optical spectroscopy \cite{75m,76m}. We
report the measurements of the degree of linear polarization for the lines
of the helium atom $\lambda =388.9$ nm for the transition $3p$ $^{3}$P$%
_{0}\rightarrow 2s$ $^{3}$S and $\lambda =587.6$ nm, the transition 3$d$ $%
^{3}$D $\rightarrow $ $2p$ $^{3}$P and nitrogen ion $\lambda =500.1-500.5$
nm, the transition 3$d$ $^{3}$F$_{0}\rightarrow 3p$ $^{3}$D due to the He$%
^{+}-$N$_{2}$ collision.

The remainder of this paper is organized as follows. In Sec. \ref{Expset}, we present
the experimental setup and measurement method. In Sec. \ref{Results}, we present the
processes and emission spectral lines measured for the He$^{+}$ $-$ N$_{2}$
and He$^{+}$ $-$ O$_{2}$ collision systems, and discuss the results of the
measurements. Finally, conclusions follow in Sec. \ref{Con}.

\section{Experiment setup and measurement method}
\label{Expset}

The radiation from the excited particles was detected using high-resolution
optical spectroscopy, which is the most precise method for the
identification of highly excited molecular states. The experimental setup
and calibration procedure are described in detail in Refs. \cite{75m,76m}.
The resolution of the optics has greatly improved since then; therefore, we
were able to distinguish the excitation channels and measure the
cross-section on the order of 10$^{-19}$ cm$^{2}$ or lower.

\begin{figure}[b]
\centering
\includegraphics[width=12.2cm]{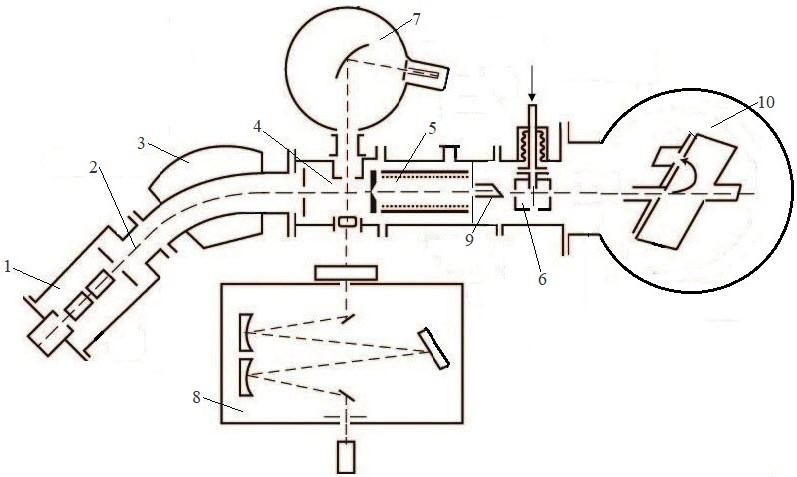}
\caption{(Color online) Schematic diagram of the experimental set up: 1 -
ion source, 2 - ion beam, 3 - magnet mass-analyzer, 4 - first collision
chamber, 5 - second collision chamber, 6 - third collision chamber, 7 -
vacuum ultraviolet spectrometer, 8 - visible spectrometer, 9 - Faraday cup,
10 - electrostatic analyzer. }
\label{F0}
\end{figure}
A schematic of the experimental setup is shown in Fig. \ref{F0}. A beam of He%
$^{+}$ ions extracted from the high-frequency (20 MHz) discharge ion source
is accelerated, collimated, \ and focused by an ion-optics system, which
includes quadruple lenses and collimating slits, and mass-selected with a 60$%
^{0}$ magnetic sector field. The beam was then directed into the collision
chamber. We used an electron gun placed in the mass-analyzer chamber to
determine the spectral sensitivity of the emission detection system. After
the collimation and additional focusing, the electron beam is directed into
the collision chamber.

The radiation emitted as a result of the excitation of colliding particles
was observed at 90$^{\text{o}}$ with respect to the direction of the ion
beam. A secondary-electron multiplier with a cooled cathode in both integral
and counting regimes detects the radiation. The spectroscopic analysis of
the emission is performed by a visible monochromator with resolution of 40
nm/mm, and by means of a Seya--Namioka vacuum monochromator with a toroidal
diffraction grating with a typical resolution of 0.05 nm that has a 1200
line/mm. The method also allowed us to measure the polarization of
excitation, which itself is a powerful tool for establishing the mechanism
for inelastic processes. A polarizer and a mica quarter-wave phase plate are
placed in front of the entrance slit of the monochromator and the linear
polarization of the emission is analyzed. For cancellation of the polarizing
effect of the monochromator, the phase plate are placed after the polarizer
and rigidly coupled to it.

The measurements at low-energy collisions required precise determination of
the energy of helium ions as well as their energy dispersion. To avoid
errors in the measurements of the energy of the incident particles, we
employed the retarding potential method and used an electrostatic analyzer
with a resolving power of 700. Additionally, by measuring the energy of the
impacting particles, we estimate the dispersion of energy provided by the
high-frequency ion source and electron gun.

The helium ion current in the collision chamber was on the order $0.1-0.5$ $%
\mu $A, while the electron current was $5-20$ $\mu $A. The pressure of the
target gas under investigation did not exceed 6$\times $10$^{-4}$Torr;
therefore, single collisions were considered. The system was differentially
pumped using an oil-free diffusion pump. The residual gas pressure did not
exceed 0.1$\times $10$^{-6}$ Torr.

The basic problem in finding the cross-section is determining the relative
and absolute spectral sensitivities of the radiation-detecting system. This
was achieved by measuring the output signal of the photomultiplier owing to
the (0.0), (0.1), (0.2), (0.3), (0.4), (1.2), (1.3), and (1.4) bands in the
first negative system of the ion N$_{2}^{+}$ (B $^{2}\Sigma _{\text{u}}^{+}-$%
X $^{2}\Sigma _{\text{g}}^{+}$ transition) and (4.0), (4.1), (6.2), (6.3),
(2.0), (3.0), (5.1), and (5.2) bands of the Meinel system (A $^{2}\Pi _{%
\text{u}}^{+}-$X $^{2}\Sigma _{\text{g}}^{+}$ transition) \cite{75m} excited
in collisions between the electrons ($E_{e}=110$ eV) and nitrogen molecules.
The output signal was normalized to the (0.1) band with the corresponding
wavelength $\lambda =427.8$ nm. This line has high intensity in this range.
The relative spectral sensitivity of the recording system obtained in this
manner was compared with the relative excitation cross-sections for the same
bands, averaged over the experimental data reported in Refs. \cite%
{48m,77m,78m,79m}. The absolute excitation cross-sections for the (0.1) band
($\lambda =427.8$ nm) were assumed to be 5.3$\times $10$^{-18}$ cm$^{2}$ at
an electron energy of 110 eV. This value was taken from Ref. \cite{14m}. The
relative and absolute uncertainties in our measurements were 5\% and 15\%,
respectively. The accuracy of the polarization measurements did not exceed $%
\sim $2 \%.

\section{ Results of experimental measurements and discussion}
\label{Results}

Sub-sections \ref{Results1} and \ref{Results2} provide an outline of the investigated processes for
the He$^{+}-$N$_{2}$ and He$^{+}$ $-$ O$_{2}$ collision systems and emission
spectral lines. In Sub-sections \ref{Results3} and \ref{Results4}, we present the experimental results and their
discussion.

\subsection{He$^{+}-$N$_{2}\ \ $collision system}
\label{Results1}

\begin{eqnarray}
\text{He}^{+}\text{(1s)}+\text{N}_{2}\text{ } &\longrightarrow &\text{ He}%
^{\ast }+\text{N}_{2}^{+}\text{ }  \notag \\
&\longrightarrow &\text{ He}^{\ast }\text{(3d}^{\text{3}}\text{D)}+\text{N}_{%
\text{2}}^{\text{+}}; \\
&\longrightarrow &\text{He}^{\ast }\text{(3p}^{\text{3}}\text{P}_{\text{0}}%
\text{)}+\text{N}_{\text{2}}^{\text{+}}; \\
&\longrightarrow &\text{He}^{\ast }\text{(4d}^{\text{3}}\text{D)}+\text{N}_{%
\text{2}}^{\text{+}}; \\
&\longrightarrow &\text{He}^{\ast }\text{(4d}^{\text{1}}\text{D)}+\text{N}_{%
\text{2}}^{\text{+}},
\end{eqnarray}%
and 
\begin{equation}
\text{He}^{+}\text{(1s)}+\text{N}_{2}\text{ }\longrightarrow \text{ He(1s}%
^{2}\text{)}+\text{N}_{2}^{+^{\ast \ast }}\text{ }
\end{equation}%
when 
\begin{eqnarray}
\text{N}_{2}^{+^{\ast \ast }}\text{ } &\longrightarrow &\text{ N}^{\ast }%
\text{(2p}^{\text{4}}\text{ }^{\text{4}}\text{P )+N}^{\text{+}}\text{(1s}^{%
\text{2}}\text{ 2s}^{\text{2}}\text{ 2p}^{\text{2}}\text{ }^{\text{3}}\text{%
P);}  \notag \\
&\longrightarrow &\text{ N}^{\ast }\text{(3s4P)}+\text{N}^{\text{+}}\text{(2p%
}^{\text{2}}\text{ }^{\text{3}}\text{P);} \\
&\longrightarrow &\text{N}^{\ast }(3\text{s}\prime \text{ }^{2}\text{D})+%
\text{N}^{\text{+}}\text{(2p}^{\text{2}}\text{ }^{\text{3}}\text{P);} \\
&\longrightarrow &\text{N}^{\ast }(4\text{p}^{2}\text{S}^{0})+\text{N}^{%
\text{+}}\text{(2p}^{\text{2}}\text{ }^{\text{3}}\text{P);} \\
&\longrightarrow &\text{N(1s}^{\text{2}}\text{2s}^{\text{2}}\text{2p}^{\text{%
3}}\text{ }^{4}\text{S}_{3/2}^{\text{0}}\text{)}+\text{N}^{\text{+*}}\text{%
(1s}^{\text{2}}\text{ 2s}^{\text{2}}\text{ 2p}^{\text{1}}\text{ 3p }^{\text{1%
}}\text{P);} \\
&\longrightarrow &\text{N(1s}^{\text{2}}\text{ 2s}^{\text{2}}\text{ 2p}^{%
\text{3}}\text{ }^{\text{4}}\text{S}_{\text{3/2}}^{\text{0}}\text{ )}+\text{N%
}^{\text{+*}}\text{(3p }^{\text{3}}\text{D);} \\
&\longrightarrow &\text{N(1s}^{\text{2}}\text{ 2s}^{\text{2}}\text{ 2p}^{%
\text{3}}\text{ }^{\text{4}}\text{S}_{\text{3/2}}^{\text{0}}\text{ )}+\text{N%
}^{\text{+*}}\text{(3d }^{\text{3}}\text{F}^{\text{0}}\text{);} \\
&\longrightarrow &\text{N(1s}^{\text{2}}\text{ 2s}^{\text{2}}\text{ 2p}^{%
\text{3}}\text{ }^{\text{4}}\text{S}_{\text{3/2}}^{\text{0}}\text{ )}+\text{N%
}^{\text{+*}}\text{(3s }^{\text{3}}\text{P}^{\text{0}}\text{);} \\
&\longrightarrow &\text{N(1s}^{\text{2}}\text{ 2s}^{\text{2}}\text{ 2p}^{%
\text{3}}\text{ }^{\text{4}}\text{S}_{\text{3/2}}^{\text{0}}\text{ )}+\text{N%
}^{\text{+*}}\text{(4f F);} \\
&\longrightarrow &\text{N(1s}^{\text{2}}\text{ 2s}^{\text{2}}\text{ 2p}^{%
\text{3}}\text{ }^{\text{4}}\text{S}_{\text{3/2}}^{\text{0}}\text{ )}+\text{N%
}^{\text{+*}}\text{(2p}^{\text{3}}\text{ }^{\text{3}}\text{D}^{\text{0}}%
\text{).}
\end{eqnarray}%
The nitrogen atom and ion lines wavelength and corresponding transitions are
the following

\begin{center}
\begin{tabular}{ccc||ccc}
\multicolumn{3}{c||}{Nitrogen atom} & \multicolumn{3}{c}{Nitrogen ion} \\ 
\hline
& Wavelength, $\lambda $ nm & Transition &  & Wavelength, $\lambda $ nm & 
Transition \\ \hline
NI & 493.5 & 4p $^{2}$S$^{0}$ $\longrightarrow $ 3s$^{\text{ \ }2}$P & NII & 
648.2 & 3p $^{1}$P $\longrightarrow $ 3s$^{\text{ \ }1}$P$^{0}$ \\ 
NI & 124.3 & 3s$^{^{\prime }}$ $^{3}$D $\longrightarrow $ 2p$^{3\text{ }4}$D$%
^{0}$ & NII & 504.5 & 3p $^{3}$S $\longrightarrow $ 3s$^{\text{ \ }3}$P$^{0}$
\\ 
NI & 120 & 3s $^{4}$P $\longrightarrow $ 2p$^{3\text{ }4}$S$^{0}$ & NII & 
500.5 & 3d $^{3}$F$^{0}$ $\longrightarrow $ 3p $^{3}$D \\ 
NI & 113.4$-$113.5 & 2p$^{4}$ $^{4}$P $\longrightarrow $ 2p$^{3\text{ }4}$S$%
^{0}$ & NII & 567.6$-$567.9 & 3p $^{3}$D $\longrightarrow $ 3s $^{3}$P$^{0}$
\\ 
&  &  & NII & 424.2 & 4f F $\longrightarrow $ 3d $^{3}$D$^{0}$ \\ 
&  &  & NII & 399.5 & 3p $^{1}$D $\longrightarrow $ 3s $^{1}$P$^{0}$ \\ 
&  &  & NII & 108.4$-$108.6 & 2p$^{3}$ $^{3}$D$^{0}$ $\longrightarrow $ 2p$%
^{2}$ $^{3}$P%
\end{tabular}
\end{center}

The helium atom lines wavelength and corresponding transitions are the
following

\begin{center}
\begin{tabular}{ccc}
\multicolumn{3}{c}{Helium atom} \\ \hline
& Wavelength, $\lambda $ nm & Transition \\ 
HeI & 667.8 & 3d $^{1}$D $\longrightarrow $ 2p$^{\text{ \ }1}$P$^{0}$ \\ 
HeI & 587.6 & 3d $^{3}$D $\longrightarrow $ 2p$^{\text{ \ }3}$P$^{0}$ \\ 
HeI & 492.2 & 4d $^{1}$D $\longrightarrow $ 2p$^{\text{ \ }1}$P$^{0}$ \\ 
HeI & 447.2 & 4d $^{3}$D $\longrightarrow $ 2p$^{\text{ \ }3}$P$^{0}$ \\ 
HeI & 388.9 & 3p $^{3}$P$^{0}$ $\longrightarrow $ 2s$^{\text{ \ }3}$S%
\end{tabular}
\end{center}

\subsection{He$^{+}$ - O$_{2}$ collision system}
\label{Results2}

\begin{eqnarray}
\text{He}^{+}\text{(1s)}+\text{O}_{2}\text{ } &\longrightarrow &\text{ He}%
^{\ast }+\text{O}_{2}^{+}\text{ }  \notag \\
&\longrightarrow &\text{ He}^{\ast }\text{(1s 2p)}+\text{O}_{\text{2}}^{%
\text{+}}; \\
&\longrightarrow &\text{He}^{\ast }\text{(3p}^{\text{3}}\text{P}_{\text{0}}%
\text{)}+\text{O}_{\text{2}}^{\text{+}}; \\
&&\text{and} \\
\text{He}^{+}\text{(1s)}+\text{O}_{2} &\longrightarrow &\text{He(2s}^{\text{2%
}}\text{)}+\text{O}_{\text{2}}^{\text{+**}}  \notag \\
&\longrightarrow &\text{He(2s}^{\text{2}}\text{)}+\text{O}^{\ast }\text{(3s}%
^{\prime }\text{ }^{\text{3}}\text{D}^{\text{0}}+\text{O}^{\text{+}}\text{(1s%
}^{\text{2}}\text{2s}^{\text{2}}\text{2p}^{\text{3}}\text{ }^{\text{4}}\text{%
S}_{\text{3/2}}^{\text{0}}\text{ );} \\
&\longrightarrow &\text{He(2s}^{\text{2}}\text{)}+\text{O}^{\text{*}}\text{%
(4d }^{\text{3}}\text{D}^{\text{0}}\text{)}+\text{O}^{\text{+}}\text{(2p}^{%
\text{3}}\ ^{\text{4}}\text{S}_{\text{3/2}}^{\text{0}}\text{ );} \\
&\longrightarrow &\text{He(2s}^{\text{2}}\text{)}+\text{O}^{\text{*}}\text{%
(3d}^{\text{3}}\text{ }^{\text{3}}\text{D}^{\text{0}}\text{)}+\text{O}^{%
\text{+}}\text{(2p}^{\text{3}}\ ^{\text{4}}\text{S}_{\text{3/2}}^{\text{0}}%
\text{ );} \\
&\longrightarrow &\text{He(2s}^{\text{2}}\text{)}+\text{O}^{\text{*}}\text{%
(3s}^{^{\prime }}\text{ }^{\text{1}}\text{D}^{\text{0}}\text{)}+\text{O}^{%
\text{+}}\text{(2p}^{\text{3}}\ ^{\text{4}}\text{S}_{\text{3/2}}^{\text{0}}%
\text{ );} \\
&\longrightarrow &\text{He(2s}^{\text{2}}\text{)}+\text{O(1s}^{\text{2}}%
\text{2s}^{\text{2}}\text{2p}^{\text{4}}\text{ }^{\text{3}}\text{P}_{\text{2}%
}\text{)}+\text{O}^{\text{+*}}\text{(1s}^{\text{2}}\text{2s}^{\text{2}}\text{%
2p}^{\text{2}}\text{3d }^{\text{3}}\text{D}^{\text{0}}\text{).}
\end{eqnarray}%
The wavelengths of the oxygen atom and ion, helium atom lines, and
corresponding transitions are as follow:

\begin{center}
\begin{tabular}{ccc||ccc}
\multicolumn{3}{c||}{Oxygen atom} & \multicolumn{3}{c}{Oxygen ion} \\ \hline
& Wavelength, $\lambda $, nm & Transition &  & Wavelength, $\lambda $, nm & 
Transition \\ \hline
OI & 99.0 & $\text{3s}^{\prime }\text{ }^{\text{3}}\text{D}^{\text{0}%
}\longrightarrow $2p$^{4}$ $^{3}$P & OII & 83.4 & 2p$^{4}$ $^{4}$P$%
\longrightarrow $2p$^{3}$ $^{4}$S$^{0}$ \\ 
OI & 97.4 & $\text{4d}\text{ }^{\text{3}}\text{D}^{\text{0}}\longrightarrow $%
2p$^{4}$ $^{3}$P &  &  &  \\ 
OI & 102.6 & $\text{3d}\text{ }^{\text{3}}\text{D}^{\text{0}}\longrightarrow 
$2p$^{4}$ $^{3}$P &  &  &  \\ 
OI & 115.2 & $\text{3s}^{^{\prime }}\text{ }^{\text{1}}\text{D}^{\text{0}%
}\longrightarrow $2p$^{4}$ $^{1}$D &  &  & 
\end{tabular}

\begin{tabular}{ccc}
\multicolumn{3}{c}{Helium atom} \\ \hline
& Wavelength, $\lambda $, nm & Transition \\ 
HeI & 53.7 & 3p $^{1}$P$^{0}\longrightarrow $1s$^{2}$ $^{1}$S \\ 
HeI & 58.4 & 2p $^{1}$P$^{0}\longrightarrow $1s$^{2}$ $^{1}$S%
\end{tabular}
\end{center}

\subsection{ Experimental results and discussion}
\label{Results3}

\begin{figure}[h]
\centering
\includegraphics[width=5.7cm]{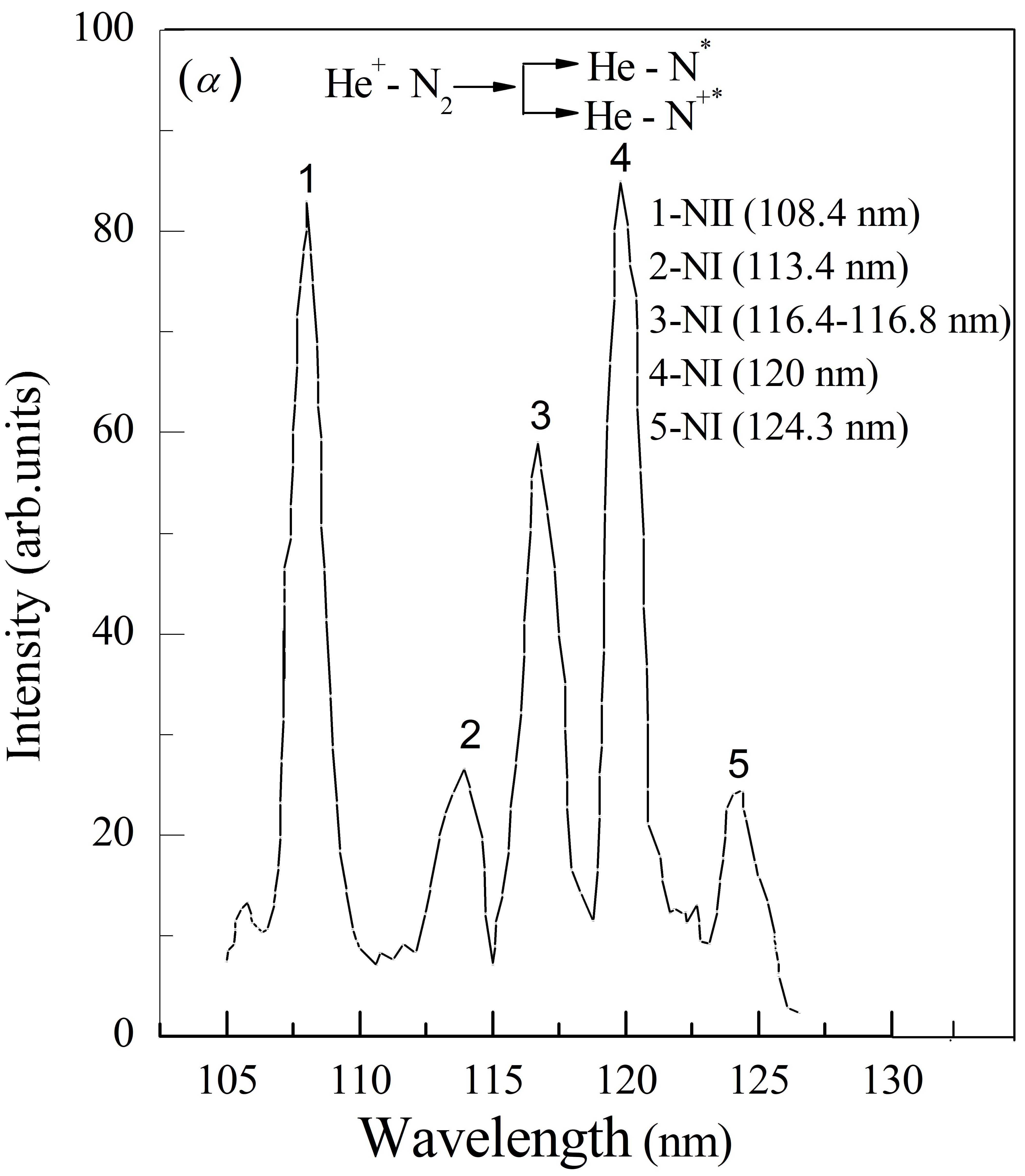} %
\includegraphics[width=7.2cm]{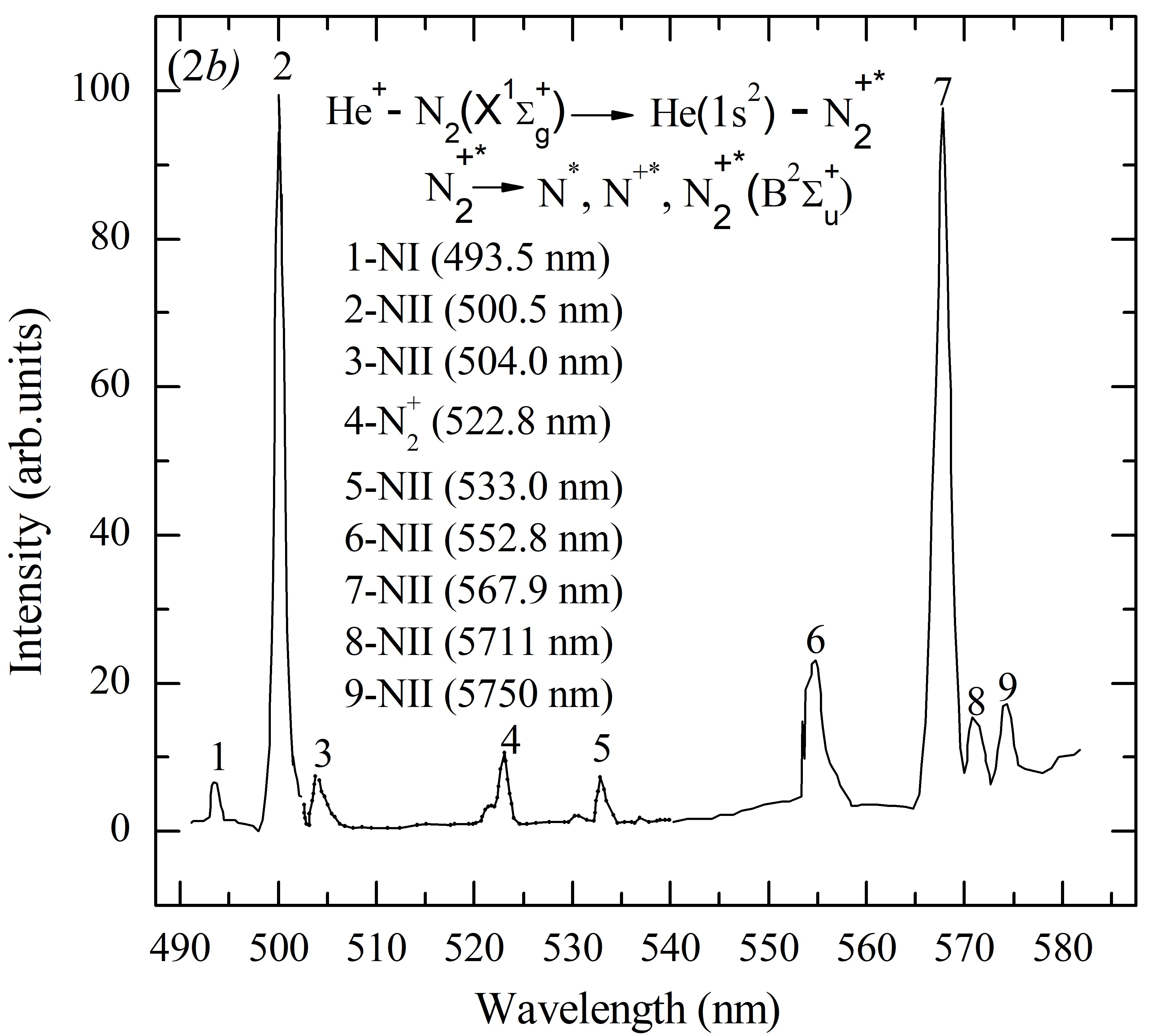}
\caption{(Color online) Dependence of the intensity on the wave length for He%
$^{+} \longrightarrow$ N$_{2}$ at ($a$) vacuum ultraviolet ($80 - 130$ nm)
and ($b$) the visible ($380 - 800$ nm) spectral regions, respectively. }
\label{F1}
\end{figure}

\begin{figure}[h]
\centering
\includegraphics[width=5.8cm]{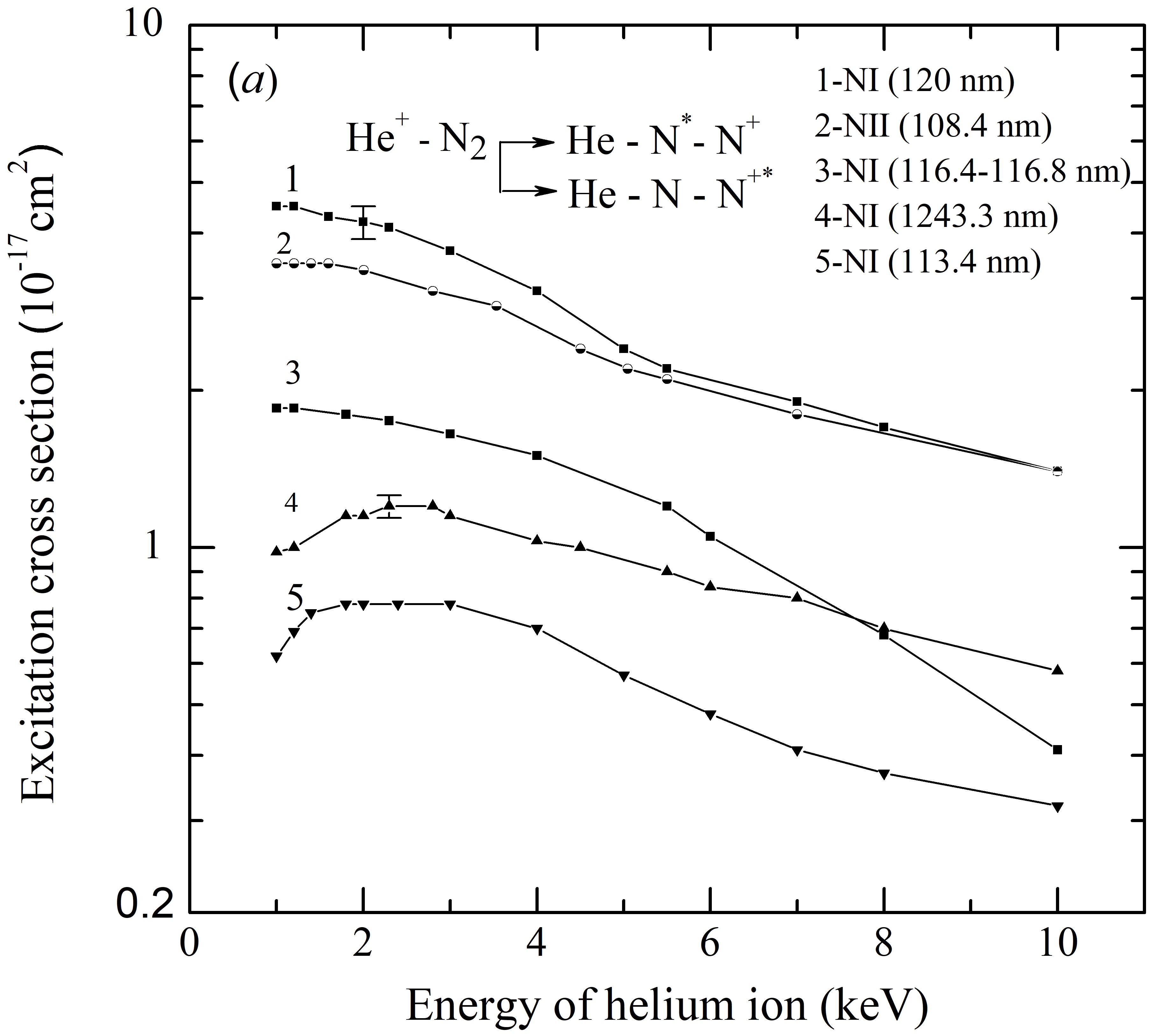} %
\includegraphics[width=5.8cm]{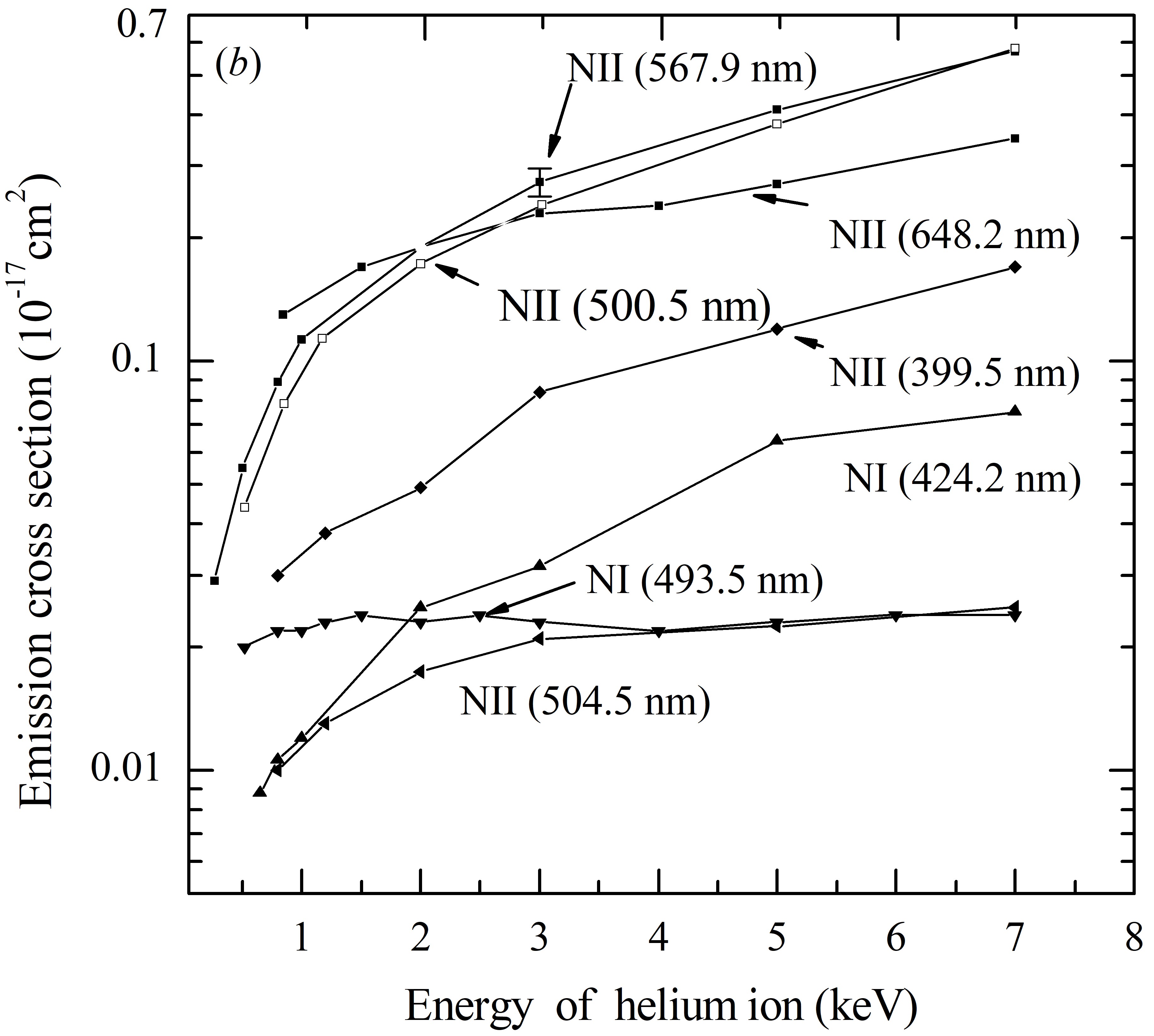} %
\includegraphics[width=5.7cm]{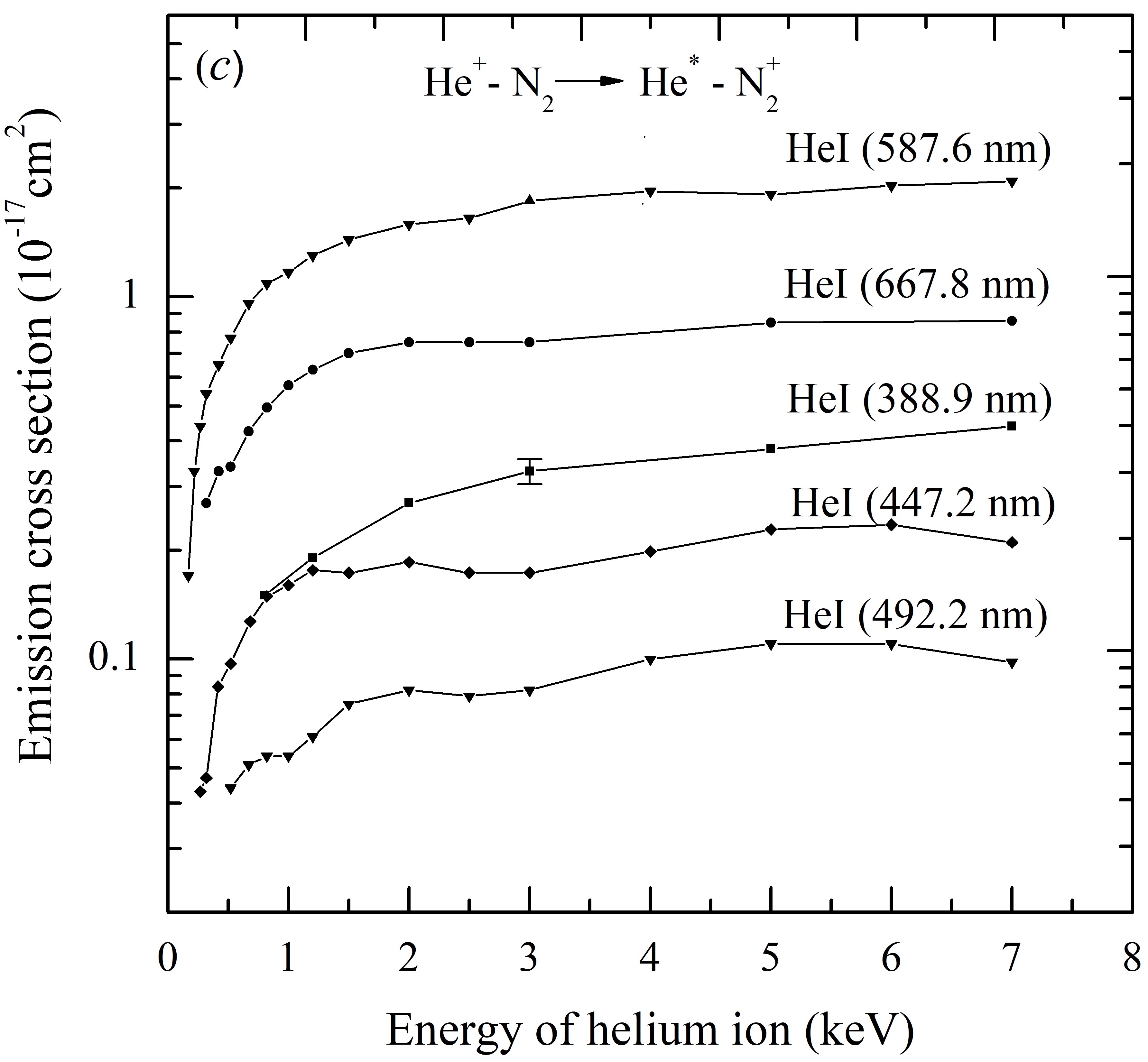}
\caption{(Color online) Energy dependence of the nitrogen atomic and ionic
lines in the He$^{+}-$ N$_{2}$ collision at ($a$) the vacuum ultraviolet and
($b$) the visible spectral regions, respectively. ($c$) Energy dependence of
the helium atom emission cross sections in the process He$^{+}-$ N$_{2}$ on
the energy of helium ions. }
\label{F3}
\end{figure}

\begin{figure}[h]
\centering
\includegraphics[width=6.2cm]{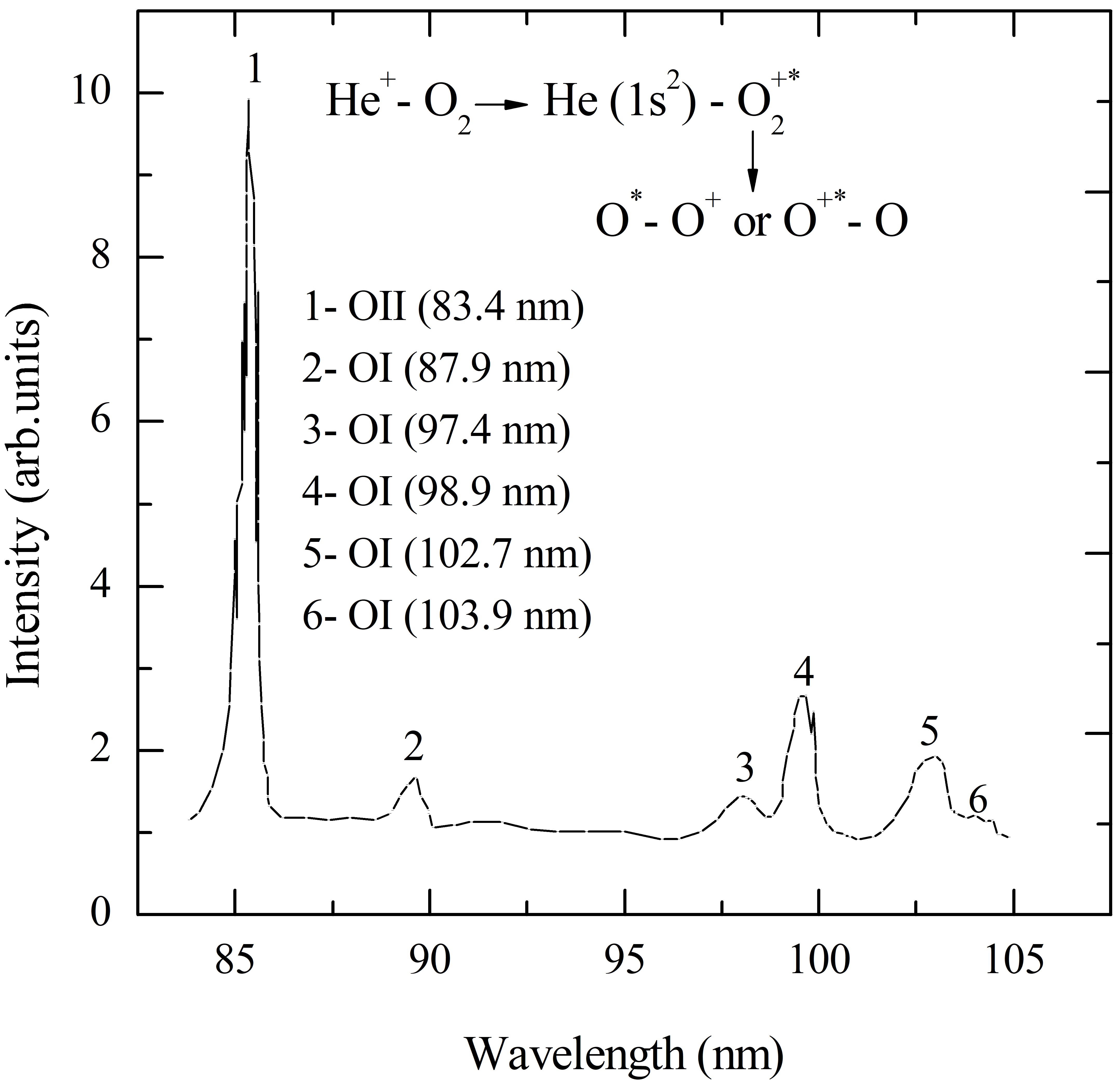}
\caption{(Color online) Emission spectrum in the wavelength range of $80-105$
nm in collisions of 10 keV helium ions with oxygen molecules. }
\label{F6}
\end{figure}

\begin{figure}[h]
\centering
\includegraphics[width=7.05cm]{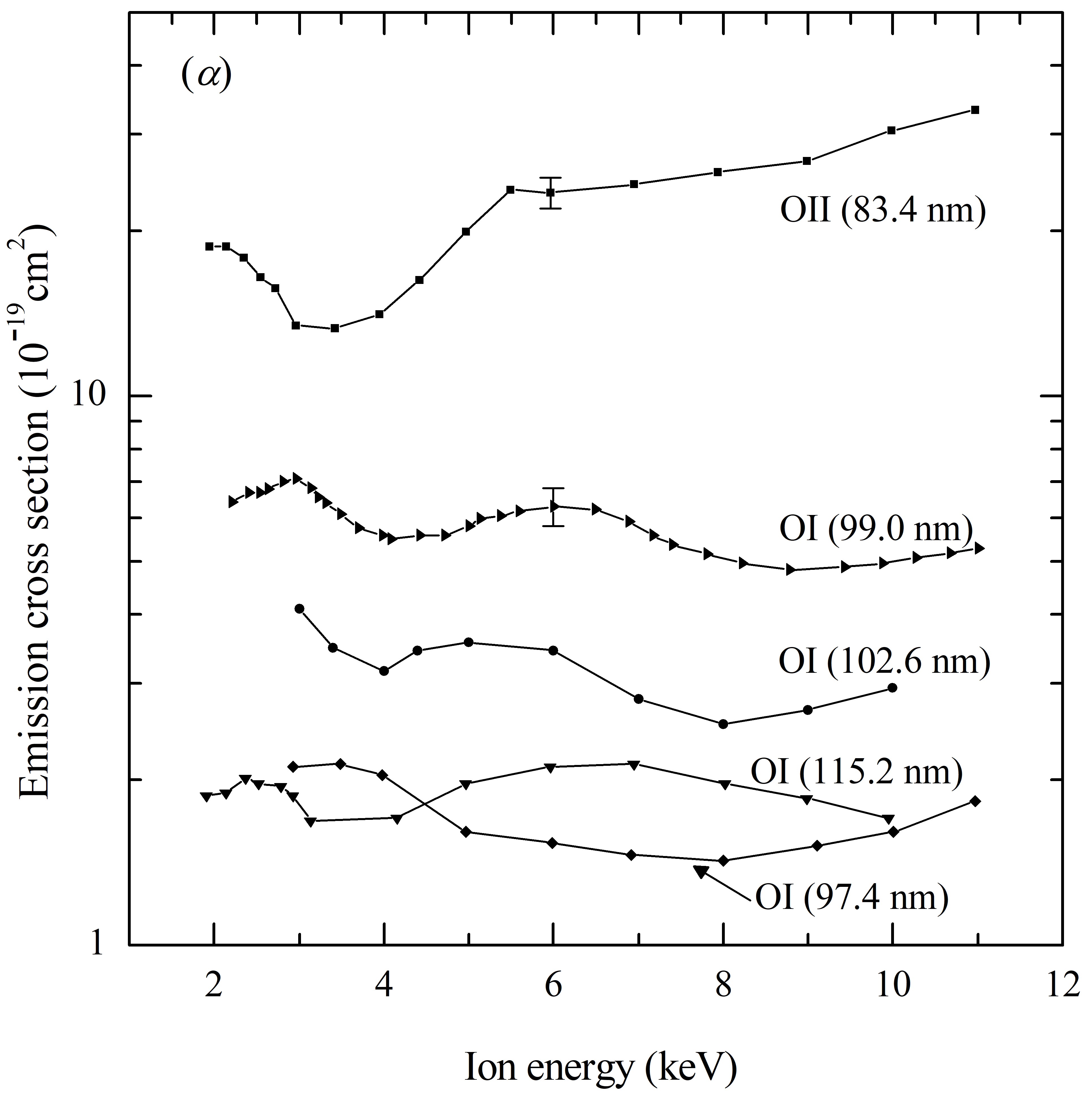} %
\includegraphics[width=7.2cm]{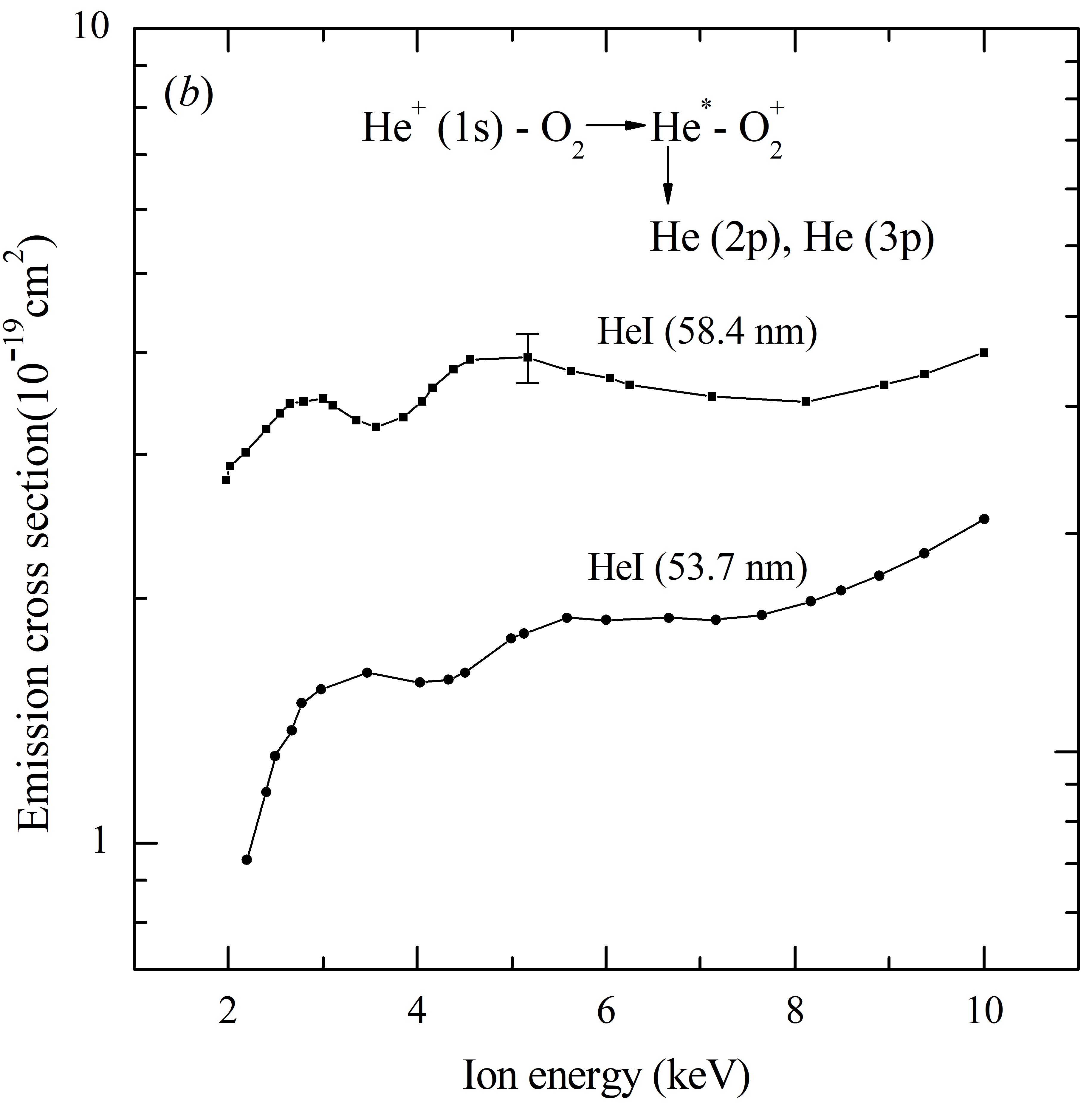}
\caption{(Color online) Energy dependences of the emission cross sections \
in He$^{+}-$O$_{2}$ collisions for ($a$) oxygen ionic OII (83.4 nm) and
atomic OI (99.0, 102.6, 115.2, 97.4 nm) lines and ($b$) for HeI atomic
(53.7, 58.4) lines.}
\label{F7}
\end{figure}

Figures \ref{F1}$a$ and \ref{F1}$b$ show the dependence of the emission
spectra on the wavelength in the vacuum ultraviolet spectral range of $%
105-130$ nm and visible spectral range of 490 -- 580 nm, respectively, for
collisions of $E=5$ keV helium ions with nitrogen molecules. Figure \ref{F1}$%
a$ presents the excitation spectra mostly for (with the exception of the
ionic line $\lambda =108.4$ nm) nitrogen atomic lines, whereas Fig. \ref{F1}$%
b$ shows nitrogen ionic lines (with the exception of the atomic nitrogen
line, $\lambda =493.5$ nm). The energy dependences of the excitation
cross-sections for the same emission spectral lines are presented in Figs. %
\ref{F3}$a$ and \ref{F3}$b$, respectively. The energy dependences of the
helium atom emission cross-sections in the He$^{+}-$ N$_{2}$ process on the
energy of helium ions are shown in Fig. \ref{F3}$c.$ The results for the
emission spectrum in the wavelength range of 80 -- 105 nm in collisions of $%
E=10$ keV helium ions with oxygen molecules are presented in Fig. \ref{F6}.
The energy dependences of the emission cross-section for oxygen atomic OI
(99.0, 102.6, 115.2, and 97.4 nm ) and oxygen ionic OII (83.4 nm) lines in He%
$^{+}-$ O$_{2}$ collisions are shown in Fig. \ref{F7}$a$.

\begin{figure}[b]
\centering
\includegraphics[width=6.0cm]{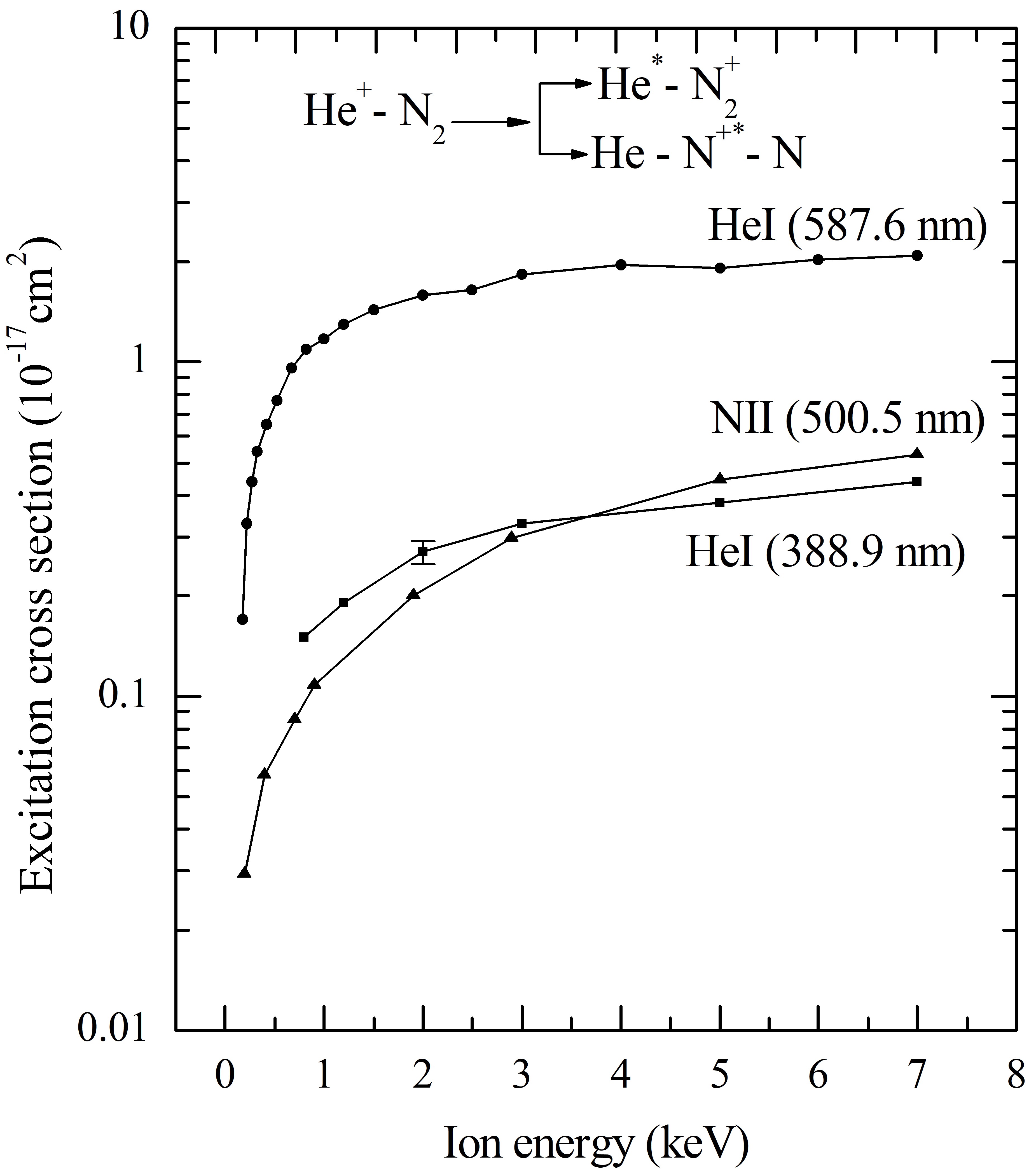}
\caption{(Color online) Energy dependence of the excitation cross section of
helium atom and nitrogen ion lines: HeI (388.9 nm) and NII (500.5 nm). }
\label{F8}
\end{figure}

From the presented measurement results, special attention is given to the
comparison of the energy dependence of helium atom excitation and nitrogen
ion excitation. For this reason, we measure the excitation functions in
visible spectral region for lines of helium atom HeI ($\lambda =$388.9 nm,
3p $^{3}$P$\longrightarrow $2s $^{3}$S) and nitrogen ion NII ($\lambda =$%
500.1 -- 500.5 nm, 3d $^{3}$F$\rightarrow 3$p $^{3}$D) as well as
polarization measurements for the same lines. The results of these
measurements are shown in Figs. \ref{F8} and \ref{F9}.

The analysis of the results shown in Figs. \ref{F1} and \ref{F6}, as well as
those shown in Figs. \ref{F3}, \ref{F7}, and \ref{F8}, and polarization
measurements presented in Fig. \ref{F9}, allows us to draw some conclusion,
related to the notable similarities of the processes considered in He$^{+}-$O%
$_{2}$ and He$^{+}-$N$_{2},$ collision systems, and the electron-impact
ionization in $e-$N$_{2}$, $e-$O$_{2}$ collision systems studied in \cite%
{IJMP2021our}.

The striking similarities in He$^{+}-$O$_{2}$ and He$^{+}-$N$_{2}$ systems
are as follow: i. the strong dominance of quasi-resonant charge-exchange
processes; ii. the dominance of endothermic processes for similar energy
defects; iii. population of exothermic channels, suggesting a strong dynamic
effect. The dominance of electron-capture processes over direct --
excitation processes, was also observed in \cite{33m}. In our case, most
similarities related to the intensity of lines, energy dependences, and
mechanisms in He$^{+}-$O$_{2}$ and He$^{+}-$N$_{2}$ collision systems are
discussed in detail below.

Let us first consider the excitation processes of the dissociation products
(nitrogen atoms and nitrogen ions) based on the results presented in Fig. %
\ref{F1} and \ref{F3}.

In Fig. \ref{F1}$a$, the curves corresponding to the two dominant atomic NI
(120.0 nm, 3s $^{4}$P$\rightarrow $2p$^{3}$ $^{4}$S) and ionic NII (108.4
nm, 2p$^{3}$ $^{3}$D$\rightarrow $ 2p$^{2}$ $^{3}$P) lines for the He$^{+}+$
N$_{2}$ collision system exhibit a similar shape. Moreover, the absolute
values of the excitation cross-sections for these lines are close to each
other. These results suggest that the excitation mechanisms of the molecular
states that dissociate into N$^{\ast }$(3s $^{4}$P) and N$^{+^{\ast }}$(2p$%
^{3}$ $^{3}$D) products are almost identical. In Ref. \cite{33m}, the
authors reached the same conclusion and showed that at relatively low
energies, $E\leq 3$ keV, charge exchange is the dominant process. According
to this study, the 1s vacancy of the He$^{+}$ plays a determining role in
different excitation processes. In the case of (HeN$_{2}$)$^{+}$ ionic
system, the initial vacancy in the He (1s) orbital becomes an inner vacancy
of the ionic atomic quasimolecule. Hence, core-excited molecular states were
formed. The formation of excited products can be related to the decay of the
intermediate molecular states of N$_{2}^{+^{\ast }}$. Specifically, the
molecular state that is correlated with either the N(3s $^{4}$P)$+$N$^{+}$%
(3P) or the N(3s $^{2}$P)$+$ N$^{+}$ ($^{3}$P) channels can be produced by
the formation of a 2s$\sigma _{g}$ hole in the ground state of the N$_{2}$
molecule \cite{80m}.

Consequently, the emission of atomic NI (120.0 nm) and ionic NII (108.4 nm)
nitrogen lines should be observed. Besides of these intense spectral lines,
we observe also atomic NI (113.4 nm; 2p$^{4}$ $^{4}$P$\rightarrow $2p$^{3}$ $%
^{4}$S$^{0}$) and ionic NII (91.6 nm; 2s2p$^{3}$ $^{3}$P$\rightarrow $2p$^{2}
$ $^{3}$p) and NII (77. 6 nm; 2s2p$^{3}$ $^{1}$D$\rightarrow $2p$^{2}$ $^{1}$%
D), which are not shown in Figs. \ref{F1}$a$ and \ref{F3}$a$, and NII (108.4
nm; 2p$^{3}$ $^{3}$D$\rightarrow $2p$^{3}$ $^{3}$P) lines (see spectral
lines in Fig. \ref{F1}$a$ and energy dependences in Fig. \ref{F3}$a$) that
are formed by the removal of a 2s electron from the inner electronic shell.
The formation of these excited products can also be caused by the decay of
core -- excited molecular states. Unfortunately, not all molecular states
that produce these lines during dissociation processes can be identified.
Therefore, the formation of highly excited molecular states has received
particular attention. These highly excited molecular states can be formed by
the removal of a 2s$\sigma _{g}$ electron from the charge-exchange channel.
To explain the excitation mechanism of this state, we used the schematic MO
correlation diagrams for the (HeN$_{2}$)$^{+}$ system from Ref. \cite{34m}.
According to this diagram, when two partners approach each other one inner 2$%
\sigma _{u}$ electron fills the He(1s) vacancy, and the other 3$\sigma _{g}$
or 1pu electrons are promoted to a high Rydberg orbital. Hence, molecular
states can be formed by using ionic cores. In these cases, a highly excited
Rydberg orbital should produce molecular states of N$_{2}^{+^{\ast }}$ with $%
^{2}\Sigma _{g}^{+}$ symmetry, which differ by two electrons from the N$_{2}$
ground state \cite{81m}. It is also possible that the formation of excited
atomic and ionic dissociation fragments can occur through decay of the $%
^{2}\Sigma _{g}^{+}$ core-excited molecular Rydberg state.

The removal of the 2s$\sigma _{g}$ electron of the N$_{2}$ molecule requires
approximately 37 eV \cite{80m}. Therefore, excitation of the inelastic
channel He(1s$^{2}$) + N$_{2}^{+}$ (2$\sigma _{g}^{-1}$) in the
charge-exchange process (the ionization potential of He is 24.6 eV) is
required to change the internal energy of the (HeN$_{2}$)$^{+}$ system by
approximately 12.4 eV. So, energy-loss spectra in the range $10.5<Q<15.6$ eV
observed in \cite{33m} in the charge exchange channel might contain this
process.

Notable similarities in the energy dependences of the He$^{+}\rightarrow $N$%
_{2}$ collision system are observed in Fig. \ref{F3}$b$. The energy
dependences and absolute values of the cross-sections are the same for
atomic ion lines of nitrogen NII (567.9 nm) with transition N$^{+}$ (3p $^{3}
$D$\rightarrow $N$^{+}$ (3s $^{3}$P$^{0}$) and NII (500.5 nm) with
transition N$^{+}$ (2p3d $^{3}$F)$\rightarrow $N$^{+}$ (2p3p $^{3}$D). The
optical excitation functions of these two lines measured for the $e-$N$_{2}$
collision system were studied in Ref. \cite{86m}, where the term excitation
function refers to the optical emission excitation function. The shapes of
these two excitation functions (5680 $\overset{o}{\text{A}}$ and 5001 $%
\overset{o}{\text{A}}$) in our study and \cite{86m} are virtually identical.
The excitation of the relatively low-intensity atomic NI (113,4 nm, 2p$^{4}$ 
$^{4}$P$\rightarrow $2p$^{3}$ $^{4}$S) and NI (124.3 nm 3s$^{I}$ $^{2}$D$%
\rightarrow $2p$^{3}$ $^{2}$D) lines presented in Fig. \ref{F3}$a$, can be
associated with the direct one- and two-electron transitions due to the MO
crossing, following MO promotion \cite{86m}.

Proceeding in the same way as in the He$^{+}-$N$_{2}$ case, we measured the
emission cross-section for the He$^{+}-$O$_{2}$ collision system. Figure \ref%
{F7}$a$ shows the energy dependence of the emission cross-section for the
atomic OI (99.0, 102.0, 115.2, and 97.4 nm) and intense ionic OII (83.4 nm)
oxygen lines. From these experimental data follows that the oxygen ion line
OII (83.4 nm, the 2p$^{4}$ $^{4}$P$\rightarrow $2p$^{3}$ $^{4}$S$_{0}$
transition) is the most intense for collisions with helium ions (Figs. \ref%
{F6} and \ref{F7}$a$). The molecular dissociation that causes excited atomic
and/or ionic fragments is due to the decay of a highly excited intermediate
molecular state of the inner shell, where a collision-induced vacancy arises.

In the case of the He$^{+}-$O$_{2}$ collision, when the particles come
closer together, a 1s inner-shell vacancy of the helium atom turns into an
inner-shell vacancy of a triatomic quasi-molecule. Accordingly, when the
particles fall apart, an unstable and highly excited O$_{2}^{+}$ molecular
ion arises. Specifically, the decay of the 2$\sigma _{g}^{-1}$vacancy in $%
^{2}\Sigma _{g}^{-}$ and \ $^{4}\Sigma _{g}^{-}$ highly excited molecular
states leads to the formation of an excited dissociation product with the
intense oxygen ion line OII (83.4 nm) \cite{82m}. 

An energy of approximately 46.2 eV is necessary to remove the 2s$\sigma _{g}$
electron from the inner shell of oxygen. Therefore, the excitation of this
inelastic channel during dissociative charge-exchange process 
\begin{equation}
\text{He}^{+}\text{(1s)}+\text{O}_{\text{2}}\rightarrow \text{He}^{+}\text{%
(1s}^{2}\text{)}+\text{O}_{\text{2}}^{+}\text{(2}\sigma _{g}^{-1}\text{)}
\end{equation}%
requires a change in the inner energy of the quasi-molecular system of (He,O$%
_{2}$)$^{+}$ roughly by 22 eV. This estimate was indirectly confirmed in
Ref. \cite{83m}. It appears that a broad peak in the energy loss spectrum
near 22 eV, typical for charge-exchange, is related to this inelastic
channel.

Let us now consider the similarities between the excitation of the He atom
and the nitrogen ion lines. Experimental data for the excitation functions
for the helium atom HeI ($\lambda =388.9$ nm, 3p $^{3}$P$\rightarrow $2s $%
^{3}$S) and nitrogen ion NII ($\lambda =500.1-500.5$ nm, 3d $^{3}$F$%
\rightarrow $3p $^{3}$D) lines are presented in Fig. \ref{F8}. The curves
exhibit surprising resemblance: in the entire investigated energy region,
both the absolute values of the emission cross-sections and their energy
dependence are close to each other. This behavior indicates the existence of
a strong correlation between the dissociation products: the excited helium
atoms and nitrogen ions

\begin{eqnarray}
\text{He}^{+}+\text{N}_{\text{2}}\text{(X }^{\text{2}}\Sigma _{g}^{+}\text{)}
&\rightarrow &\text{He}^{\ast }+\text{N}^{+}+\text{N,}  \label{He12} \\
\text{He}^{+}+\text{N}_{\text{2}}\text{(X }^{\text{2}}\Sigma _{g}^{+}\text{)}
&\rightarrow &\text{He}+\text{N}_{2}^{+^{\ast }}\rightarrow \text{He}+\text{N%
}^{+^{\ast }}+\text{N.}  \label{He13}
\end{eqnarray}%
In addition, we assume that the inelastic energy defects for these channels
are close to each other. Some additional arguments substantiate the
existence of a correlation between channels (\ref{He12}) and (\ref{He13}).
In Ref. \cite{86m} the electron-impact dissociative excitation of nitrogen
molecules N$_{2}$ was investigated. The authors observed the same emission
line for N$^{+}:$ \ $\lambda =500.5$ nm, corresponding to the transition 3d $%
^{3}$F$\rightarrow $3p $^{3}$D. Because the incident particle electron has a
small mass, the experimentally obtained threshold energy of 57 eV, for the
appearance of this line nearly coincides with the corresponding energy
defect for this process. Therefore, for the threshold of 57 eV, after
reduction by the ionization potential of the helium atom (24.6 eV), gives
approximately 32 eV for the energy defect. In the energy loss spectrum
plotted in \cite{33m}, for the charge-exchange channel, one can find a broad
peak in the vicinity of $\sim $30 eV. Thus, 32 eV is located in this area,
which is indirect evidence in favor of the close relationship between
reactions (\ref{He12}) and (\ref{He13}) \cite{33m}.

\subsection{Polarization}
\label{Results4}

Let us consider the polarization following Ref. \cite{Malxaz}. The
polarization of the emission emerging from the excited $^{3}$P-state of
helium is related to the relative populations of $m_{L}=0$ and $m_{L}=\pm 1$
sublevels. The expression for the first Stokes' parameter is derived based
on the general approach developed in \cite{35m}. The Appendix of Ref. \cite%
{Malxaz} presents the details of these calculations, and the final formula
for the linear polarization is as follows:

\begin{equation}
P=\frac{\mathfrak{I}_{\parallel }-\mathfrak{I}_{\perp }}{\mathfrak{I}%
_{\parallel }+\mathfrak{I}_{\perp }}=\frac{15\left( \sigma _{0}-\sigma
_{1}\right) }{41\sigma _{0}+67\sigma _{1}},  \label{P11}
\end{equation}%
where $\mathfrak{I}_{\parallel }$ and $\mathfrak{I}_{\perp }$ are the
intensities of the radiation emitted in a direction perpendicular to the
helium beam having electric vectors parallel and perpendicular to the beam
direction, respectively. In Eq. (\ref{P11}), $\sigma _{0}$ and $\sigma _{1}$
represent cross-sections for\ the population of sublevels with $m_{L}=0$ and 
$m_{L}=\pm 1$, respectively. Our experimental observation leads to a value
of $P\sim 20$\% in the energy range of $6.5-10$ keV. From Eq. (\ref{P11}),
we obtained the ratio $\frac{\sigma _{0}}{\sigma _{1}}\approx 15.$ Such a
large ratio value indicates that $m_{L}=\pm 1$\ sublevels of the excited
helium atom are preferably populated. The latter implies that the electron
density formed in the He$^{\ast }$ during the collision is oriented
perpendicularly with respect to the incident beam direction.  
\begin{figure}[h]
\centering
\includegraphics[width=6.0cm]{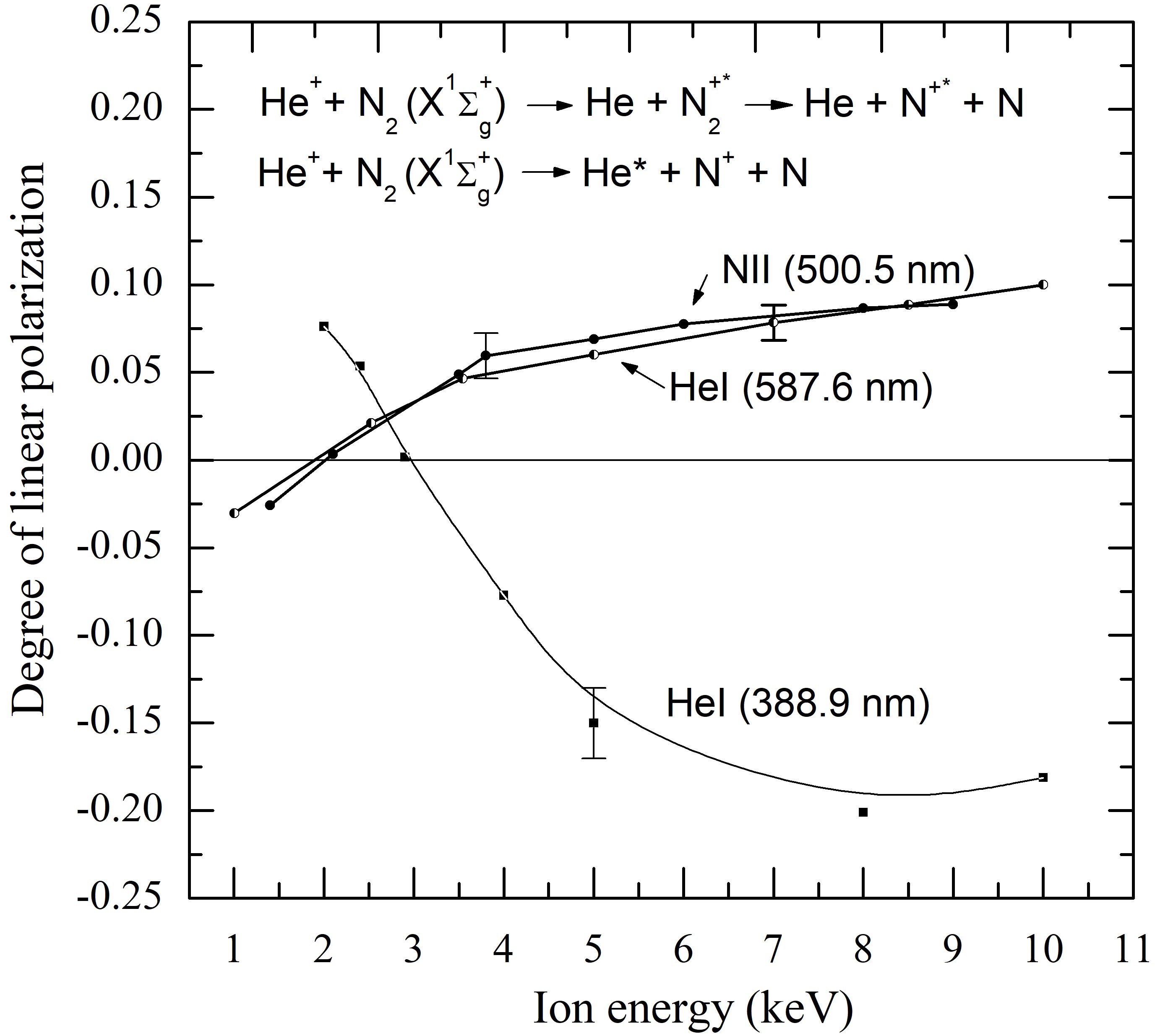}
\caption{(Color online) Energy dependence of the degree of polarization: HeI
(388.9 nm), NII (500.5 nm), \ HeI (587,6 nm), our error bars represent one
standart deviation. }
\label{F9}
\end{figure}
In Fig. \ref{F9}, the results of the polarization measurements are
presented. The results indicate the difference and similarities of
polarization for the investigated emission lines. As shown, the maximum
negative degree of polarization is 20\% at the energy 8 keV for the HeI
(388.9 nm) line, which is $\sim $ 6\% for the NII ($500.1-500.5$ nm), which
is a dissociation product. As shown in Fig. \ref{F9}, the degrees of
polarization for the HeI (587.6 nm) \ and NII (500.5 nm) emission lines
change the sign at nearly the same energy $\sim $ 2.3 keV and appear to be
independent of the He$^{+}$ incident energy in the range of $5-10$ keV for
the N$^{+}$ and 8 to 10 keV for He. A rise in polarization as the energy
decreases was also been noted for this transition in \cite{84m,85m}. They
found that the polarization decreased to a value of approximately 5\%. This
result is consistent with our results. For the polarization of radiation
emitted by the nitrogen ion N$^{+}$ (dissociation product) and helium atom He%
$^{\ast }$(3d $^{3}$D) we use the same technique as in Ref. \cite{Malxaz}.

The energy dependence of the measured polarizations showed that the
electronic orientation of the excited He atom changed at nearly 3 keV. It
can be assumed that, because of the strong correlation between the
excitation channels of He and N$^{+},$ the electronic orientation of the
excited nitrogen ion would also change. This implies that the effect of the
molecular axis orientation with respect to the incident ion beam also
changes as the energy increases.

\section{Conclusions}
\label{Con}

The striking similarities between the processes realized in the He$^{+}+$N$%
_{2}$ and He$^{+}+$O$_{2}$ collision systems have been reported. In
collisions of helium ions with oxygen and nitrogen molecules, the intense
atomic and ionic lines obtained are largely related to charge-exchange
processes \cite{13m}. In both cases, the excited dissociative products
(oxygen ion/atom and nitrogen ion/atom) are formed through the decay of the
highly excited molecular states of the O$_{2}^{+^{\ast }}$ and N$%
_{2}^{+^{\ast }}$, respectively.

We observed a similar shape of energy dependence, almost the same value of
excitation cross-sections, as well as a common excitation mechanism for two
dominant dissociative nitrogen atomic N$^{\ast }$(3s $^{4}$P) and nitrogen
ionic N$^{+^{\ast }}$(2p$^{3}$ $^{3}$D) products in the He$^{+}+$N$_{2}$
collision. We found that these excited products could be formed by the
removal of a 2s$\sigma _{g}$ electron in the charge-exchange channel.
Notable similarities in the energy dependences and absolute values of the
cross-sections for the nitrogen ionic lines NII (567.9 nm) and NII (500.5 nm
) in He$^{+}+$N$_{2}$ collisions were observed. The strong correlation
between the excitation of the helium atomic HeI (388.9 nm, transition 3d $%
^{3}$F$\longrightarrow $ 3p $^{3}$D) and the nitrogen ionic NII ($500.1-500.5
$ nm, transition 3d $^{3}$F$\rightarrow $3p $^{3}$D) lines are revealed for
the He$^{+}+$N$_{2}$ collision system. The most intense oxygen ionic OII
(83.4 nm) and atomic OI (99.0 nm) lines and weak (approximately 10$^{-19}$ cm%
$^{2}$), double charged oxygen ion OIII (70.6 nm) lines were observed and
identified in the collision of He$^{+}$ with O$_{2}$ molecules. In this
case, the molecular dissociation that causes the excited atomic and/or ionic
fragments is due to the decay of a highly excited intermediate molecular
state of the inner shell, where a collision-induced vacancy arises. The
highly excited molecular states ($^{2}\Sigma _{g}^{-}$ and $^{4}\Sigma
_{g}^{-}$) in He$^{+}+$O$_{2}$ collisions are assigned, and their leading
roles in explaining the mechanism of the intense oxygen ionic OII (83.4 nm)
and atomic OI (99.0 nm) lines are explained. Energy dependence of the degree
of linear polarization for the He atomic lines HeI (388.9 nm and 587.6 nm)
and nitrogen ionic line NII (500.5 nm) are measured. The maximum negative
(20\%) and minimum (5\%) positive values of the degree of linear
polarization are revealed for dissociative products of the helium atom
(388.9 nm) and helium atom and nitrogen ion (587.6 nm; 500.1 nm),
respectively, at the same collision energy $E=2.5$ keV. Based on
polarization measurements, the cross-section $\sigma _{0}$ and $\sigma _{1}$%
, related to the relative population of the helium He$^{\ast }$ (3P)
magnetic sublevels with $m_{L}=0$ and $m_{L}=\pm 1,$ respectively, are
calculated, and the ratio $\sigma _{1}/\sigma _{0}\approx $15 is revealed.
Such a high value of the ratio indicates that: i. the sublevels of the
excited helium $^{3}$P state are preferably populated; ii. the electron
density formed in the excited helium He$^{\ast }$ atom during the collision
was oriented perpendicular to the incident beam direction. Most of the
experimental data obtained for the He$^{+}+$N$_{2}$ and He$^{+}+$O$_{2}$
collision systems were qualitatively interpreted in terms of quasi-diatomic
approximation.

\end{document}